\documentclass[twocolumn,showpacs,nofootinbib,10pt]{revtex4-1}

\usepackage{color}
\usepackage{graphicx}
\usepackage{graphicx,float}\usepackage[all]{xy}
\usepackage{amsmath,upgreek}
\usepackage{amssymb,bbm}

\newcommand{\beq}{\begin{eqnarray}}
\newcommand{\eeq}{\end{eqnarray}}
\newcommand{\be}{\begin{equation}}
\newcommand{\ee}{\end{equation}}
\newcommand{\bea}{\begin{eqnarray}}
\newcommand{\eea}{\end{eqnarray}}

\newcommand{\ba}{\begin{eqnarray}}
\newcommand{\ea}{\end{eqnarray}}
\newcommand{\approptoinn}[2]{\mathrel{\vcenter{
  \offinterlineskip\halign{\hfil$##$\cr
    #1\propto\cr\noalign{\kern2pt}#1\sim\cr\noalign{\kern-2pt}}}}}

\newcommand{\appropto}{\mathpalette\approptoinn\relax}
\usepackage[colorlinks,hyperindex]{hyperref}

\begin{document}
\title{
Quadratic curvature terms  and deformed Schwarzschild-de Sitter black hole analogues in the laboratory}

\author{Rold\~ao da Rocha}
\affiliation{Centro de Matem\'atica, Computa\c c\~ao e Cogni\c c\~ao, Universidade Federal do ABC - UFABC\\ 09210-580, Santo Andr\'e, Brazil.}
\email{roldao.rocha@ufabc.edu.br}

\author{R. F. Sobreiro}
\affiliation{UFF - Universidade Federal Fluminense,
Instituto de F\'isica, Campus da Praia Vermelha,
Avenida General Milton Tavares de Souza s/n, 24210-346,
Niter\'oi, RJ, Brazil.}
\email{sobreiro@if.uff.br}

\author{A. A. Tomaz}
\affiliation{CBPF -- Centro Brasileiro de Pesquisas F\'isicas,
Rua Dr. Xavier Sigaud, 150\\ 22290-180
Rio de Janeiro, RJ, Brazil}
\email{tomaz@cbpf.br}

%\date{December 25, 2016}

\begin{abstract}

Sound waves on a fluid stream, in a de Laval nozzle, are shown to correspond to
 quasinormal modes
emitted by black holes that are physical solutions in a quadratic curvature gravity with cosmological constant. Sound waves patterns in transsonic regimes at a laboratory are employed here to provide experimental  
data regarding generalized
theories of gravity, comprised by the exact de Sitter-like
solution and a perturbative
solution around the Schwarzschild-de Sitter standard solution. Using the classical tests of General  Relativity  to bound free parameters in these solutions, acoustic perturbations
on fluid flows in nozzles are then regarded to study quasinormal modes
of these black holes solutions, providing deviations of the de Laval nozzle cross-sectional area, when compared to the Schwarzschild solution. The fluid sonic point  in the nozzle, for sound waves in the fluid, implements the acoustic event horizon corresponding to quasinormal modes.

 \end{abstract}



\maketitle

\section{Introduction} 

Perturbing the underlying geometry surrounding black holes yields the emission of a peculiar ringing wave pattern, namely, a quasinormal ringing \cite{Kokkotas:1999bd,Nollert:1999ji,Konoplya:2011qq}.  
Quasinormal modes are physical signatures that can reveal the very nature of, for example, black hole mergers, playing relevant roles on the observation of gravitational waves radiation emitted 
by such kind of systems. Within this setup, black hole physics can be scrutinized by studying fluid dynamics. Indeed, sound waves can be studied in the propagation of gas flows. The sonic point, defined by fluid flows that reach the speed of sound, plays the role of an event horizon for sound waves. Instead, supersonic regions, wherein the flow has speeds higher than the speed of sound, correspond to the black hole inner region. In this scenario, transonic flows are correlated to the so called acoustic black holes \cite{Visser:1997ux} that are, thus, utilized to test black holes quasinormal modes in, for instance, a laboratory of propulsion. 

In fluid flows, sonic points generate a surface that plays the role of an event horizon, being realized by the sound waves as an acoustic horizon, or, surface gravity. In this setup, sound waves can be perturbed, producing 
frequency modes that are similar to the frequency of gravitational wave radiation  emitted by black holes mergers  \cite{Abdalla:2007dz,Cuyubamba:2013iwa,Kokkotas:1999bd,Nollert:1999ji,Santos:2015gja}.
The  surface gravity related to (acoustic) black holes has been already produced in laboratories  \cite{Furuhashi:2006dh}, perturbing fluid flows in a nozzle. Such kind of experiments implements, correspondingly, perturbations of  black holes analogues. The prototypical Schwarzschild black hole has been studied in this context \cite{Okuzumi:2007hf,Abdalla:2007dz,Cuyubamba:2013iwa}, as well as black holes on fluid brane-world models 
\cite{daRocha:2017lqj}. 

 A de Laval nozzle is a propelling nozzle that accelerates  pressurized gas streams at high temperatures into either transonic or supersonic (or even hypersonic) speeds. 
 de Laval nozzles are built to essentially make the fluid flow to taper down, up to the pinch point, and then to flare outwards  the divergent cusp in the nozzle. The outgoing fluid flow can be shot out at supersonic or hypersonic rates.
Any stable transonic flow can be implemented in laboratories, when the converging part of the nozzle 
has a different pressure than the one at the divergent nozzle cusp \cite{landau1987fluid}. 
In fact, the de Laval nozzle tapers down at the converging cusp, forcing the fluid flow to go faster until it reaches the speed of sound. At such speed, a divergent cusp makes the pressure on the fluid to increase. As the stream narrows, its speed increases and the fluid mass flow rate remains constant, until the fluid flow reaches the speed of sound. This position and regime define the choke point, where the fluid does not flow %-- across the convergent cusp at the nozzle -- 
any faster, even if the nozzle gets narrower. After the choke point into the divergent cusp, flaring the nozzle out again makes the pressure on the fluid flow at the choke point to dissipate and the fluid accelerates faster than the speed of sound. Throughout this process, the thermal energy of the fluid flow is swapped into kinetic energy, causing the fluid speed to increase to speeds higher that the speed of sound. Quasi-1D fluid flows underlie those kind of models, where the density of the fluid can vary and the flow can be compressed \cite{landau1987fluid}.

It is well known for all physicists the accuracy of the General Theory of Relativity (GR) proposed by Einstein, in dealing with long lengths. On the other hand, extensions and/or modifications of GR is in vogue since, for instance, the \emph{adventum} of gauge theories of gravitation, to reach domains where GR are not well-succeeded. A novel induced gravity theory \cite{Sobreiro:2011hb} was used to derive a perturbative solution around a Schwarzschild-de Sitter (SdS) geometry \cite{Silveira:2017ijo}. The core of such solution concentrates on understanding the influence of a quadratic curvature term in the field equations, even in a torsionless setup. The arbitrariness investigated in Ref. \cite{Silveira:2017ijo} goes towards the implementation of perturbation methods to solve a differential equation extracted from every modified theory of gravity that sustains any contribution of a quadratic curvature. Furthermore, all kind of perturbative solutions in \cite{Silveira:2017ijo} can open doors towards investigations of the first simplified SdS black holes and their respective contributions to study their emission of gravitational waves from an astrophysical point of view. This study shall be here implemented through 
the duality between quasinormal modes emitted from such kind of black holes mergers and sound waves perturbations in a de Laval nozzle.

This paper is organized as follows:  Sect. II is devoted to implement  physical black hole solutions of the equations that are derived from the gravitational action involving a quadratic curvature term and a cosmological constant. 
Sect. III studies quasi-1D, adiabatic, isentropic fluid flows and their perturbations in a de Laval nozzle. The associated  wave equation  is shown to be equal to the wave equation for   perturbations of spin-$s$ type, regarding black holes that are solutions of the action with  a quadratic curvature term. Quasinormal modes emitted from such kind of black holes mergers can be then analysed by studying the equivalent  wave equations, once the de Laval cross-sectional nozzle coordinate is expressed  as a function of the black hole radial coordinate. We explicitly study the case for $s=\ell=0,1$, 
showing how the quadratic curvature corrections for the Einstein-Hilbert action induce modifications into the corresponding de Laval nozzle cross-section, and their consequences. Sect. IV regards the conclusion,  discussions and analysis of the previous results.

\vspace*{-0.5cm}
\section{The $R^2$ setup}

This section is devoted to briefly present the static, spherically symmetric solution, derived in Ref. \cite{Silveira:2017ijo}, involving a quadratic curvature term. 
\subsection{Action and field equations}\label{odesys}

The theory of gravity considered in Ref. \cite{Sobreiro:2011hb} is represented by the following gravitational action,
\begin{eqnarray}\label{ym-map-grav-obs}
\!\!\!\!\!S_{\mathrm{grav}}&\!=\!&\!\frac{1}{16\pi G}\!\int\! \bigg(\frac{3}{2\Lambda^2}R^{\mathfrak{a}}_{~\mathfrak{b}}\star R_{~\mathfrak{a}}^{\mathfrak{b}} \!- R_\mathfrak{ab}\star (e^\mathfrak{a}{} e^\mathfrak{b})\!+\! T_\mathfrak{a}\star T^\mathfrak{a}\nonumber\\&&\qquad\qquad\qquad\qquad\quad + \frac{\tilde{\Lambda}^2}{6}[\star (e_\mathfrak{a}{} e_\mathfrak{b})] e^\mathfrak{a}{} e^\mathfrak{b}\bigg)~.
\end{eqnarray}
with $R_\mathfrak{a}^{~\mathfrak{b}} = d \omega_\mathfrak{a}^{~\mathfrak{b}}+\omega_{\mathfrak{a}}^{~\mathfrak{c}} \omega_\mathfrak{c}^{~\mathfrak{b}}$ standing for the curvature 2-form and $T^\mathfrak{a}=d e^\mathfrak{a}+\omega^\mathfrak{a}_{~\mathfrak{c}}e^\mathfrak{c}$ for the torsion 2-form. The 1-form spin connection is represented by $\omega^\mathfrak{a}_{~\mathfrak{b}}$  and $e^\mathfrak{a}$ displays the vierbein 1-form. Furthermore, gothic indexes $\{\mathfrak{a},\mathfrak{b},\mathfrak{c},\mathfrak{d}\}$ run as $\{0,1,2,3\}$. The Hodge dual operator is denoted by $\star$, whereas $G$ stands for  the Newton's constant, $\tilde{\Lambda}^2$ is the cosmological constant and $\Lambda^2$ is a mass parameter. Using the action \eqref{ym-map-grav-obs} yields the field equations
\begin{subequations}
\begin{eqnarray} 
\hspace*{-2cm}\!\!\!\!\!\!\!\!\!\!\!\!\!\!\!\!\!\!\!\frac{3}{2\Lambda^2}R_{\mathfrak{bc}} \star (R^{\mathfrak{bc}}e_{\mathfrak{a}})+D\star T_{\mathfrak{a}}+ T_{\mathfrak{b}}\star\left(T^{\mathfrak{b}}e_{\mathfrak{a}}\right)\hspace*{0.59cm}&&\nonumber\\- \varepsilon_{\mathfrak{abcd}}R^{\mathfrak{bc}}e^{\mathfrak{d}} - \tilde{\Lambda}^{2}\star e_{\mathfrak{a}} &=& 0~,\label{eq:eq_mov1}\\
\!\!\!\!\varepsilon_{\mathfrak{abcd}} T^{\mathfrak{c}} e^{\mathfrak{d}}+e_{\mathfrak{a}} \star T_{\mathfrak{b}}-\frac{3}{\Lambda^2}D\star R_{\mathfrak{a}\mathfrak{b}} 
 -e_{\mathfrak{b}} \star T_{\mathfrak{a}}
 &=& 0~,\label{eq:eq_mov2}
\end{eqnarray}
\end{subequations}
for the spin connection and for the vierbein fields, respectively.

The first task consists to investigate the most simple scenario, since the coupled system (\ref{eq:eq_mov1} - \ref{eq:eq_mov2}) is a non-trivial one to straightforwardly extract any solution. Therefore, Eqs.~\eqref{eq:eq_mov1} and~\eqref{eq:eq_mov2} can be simplified by setting the torsion equal to zero and by  multiplying the equations by $\lambda\equiv-\tilde{\Lambda}^2/3$. The ratio $\zeta\equiv\Lambda^2/(2\tilde{\Lambda}^2)$ regards a small parameter, since originally, $\Lambda^2\gg\tilde{\Lambda}^2$, at a minimum.  Numerically manipulating the ratio $\zeta$  makes us 
to understand the influence of the quadratic curvature term in a perturbative way to solve the remaining differential equation \cite{Silveira:2017ijo}. The parameter $\zeta$ enforces Eq.~\eqref{eq:eq_mov2} not to have a perturbative solution, since it is the perturbed  term itself. Hence, the second equation can be discarded. More details  can be seen in Ref. \cite{Silveira:2017ijo}, wherein the following equation of motion 
was derived:
\begin{equation}\label{eq:eq_mov1mod}
 \varepsilon_{\mathfrak{abcd}}R^{\mathfrak{bc}} e^{\mathfrak{d}}-\zeta R_{\mathfrak{bc}} \star (R^{\mathfrak{bc}}e_{\mathfrak{a}})  - 3\lambda \star e_{\mathfrak{a}} = 0~.
\end{equation}
Using Schwarzschild coordinates
\begin{equation}\label{spheric-symm}
\!\!\!\!e^0=e^{\tau(r)}\mathrm{d}t,~~e^1=e^{\beta(r)}\mathrm{d}r,~~e^2=r\mathrm{d}\theta,~e^3=r \sin\theta \mathrm{d}\varphi~
\end{equation}
in Eq.~\eqref{eq:eq_mov1mod}, adopting the standard choice  $\tau=-\beta$   makes the equation for $\mathfrak{a}=0$ to be equal to the one for $\mathfrak{a}=1$. Besides, the differential equations obtained for $\mathfrak{a}=0$ and $\mathfrak{a}=2$ are equivalent. Hence, they can be independently solved by perturbation methods \cite{kelley2010theory}. Therefore, the only equation to be solved  perturbatively reads \begin{eqnarray}\label{edo-t-pert}
&&\!\!\!\!\!\!\!\!\!\!\!\!\!\!\!\!\!\!\!\!\!\!\!\frac{\zeta}{r^4}\left[\left(1\!-\!e^{-2\beta}\right)^2\!+\!{2r^2}\left(e^{-2\beta}\partial_r\beta\right)^2\right]\nonumber\\&&+\frac{2\lambda}{r^2}\left[3\lambda r^2+\!1\!-\!e^{-2\beta}+r{e^{-2\beta}\partial_r\beta}\right] =0~.
\end{eqnarray}

\subsection{Perturbative solution}\label{pertsol}

From the perturbation theory, we consider the quadratic curvature as a small perturbation in Eq.~\eqref{eq:eq_mov1mod}. Let $u(r)=1-e^{-2\beta}$ and rewrite Eq.~\eqref{edo-t-pert} as
\begin{equation}\label{edo-t-u}
\zeta\left[\frac{1}{2}{\dot{u}}^2+\frac{u^2}{r^2}\right] + \lambda\left(3\lambda r^2+u+r\dot{u} \right)=0~,
\end{equation}
where $\dot{u}\equiv du(r)/dr$. Clearly, all derivatives are ordinary, since $\beta\equiv \beta(r)$ is only $r$-dependent. 
A perturbative solution of Eq.~\eqref{edo-t-u} requires a general expression as
\begin{equation}\label{SPert-geral}
u(r)=u_0(r) + {\sum_{i=1}^\infty\zeta^i u_i(r).}
\end{equation}
Replacing \eqref{SPert-geral} in Eq.~\eqref{edo-t-u} yields, for each order \footnote{As a matter of clarity, here the dimensionless parameter is renamed by $\zeta$, instead of the $\eta$ in Ref.\cite{Silveira:2017ijo}.} in $\zeta$, an infinite set of hierarchical equations
 {\begin{eqnarray}\label{hierarq-pert-u}
r \dot{u_0}+u_0+3\lambda r^2 &=& 0~,\nonumber\\
\lambda (r\dot{u_1}+ u_1) + \frac{1}{2}\dot{u}_0^2 + \frac{u_0^2}{r^2} &=& 0~,\nonumber\\
\lambda (r\dot{u}_2+ u_2)+\dot{u}_1\dot{u}_0 + \frac{2u_0 u_1}{r^2} &=& 0~,\nonumber\\
\vdots\qquad\qquad&=&\;\vdots
\nonumber\\
\ldots\qquad\qquad&=&0\;.
\end{eqnarray}}
Solving the above system iteratively implies that 
\begin{equation}
u_0 = \frac{\tilde{\Lambda}^2r^2}{3}  + \frac{2GM}{r}~,
\end{equation}
as the $0^{\rm th}$ order solution, corresponding to the usual Schwarzschild-de Sitter solution \cite{Gibbons:1977mu,Cardoso:2003sw}. The integration constant is derived by regarding the Newtonian limit. Subsequently, the first order solution can computed, 
\begin{equation}\label{pert-sol-1th}
\!\!\!\!e^{-2\beta} \!\approx\! 1\!-\!\frac{2GM}{r}\!-\!\frac{\tilde{\Lambda}^2}{3}r^2\!-\!\zeta\!\left(\frac{\mathcal{C}_{12}}{r}\!+\!\frac{\tilde{\Lambda}^2}{3}r^2\!+\!\frac{6G^2M^2}{\tilde{\Lambda}^2 r^4}\right).
\end{equation}
It is worth to realize that the limit $r\gg 2GM$ provides a typical perturbative solution around a de Sitter spacetime. In the next section, we shall use  \eqref{pert-sol-1th} 
to derive the corresponding de Laval nozzle profile ruled by the solution \eqref{pert-sol-1th}. 

\section{de Laval nozzle in the quadratic curvature setup}

 de Laval nozzles are constructed upon the theory of quasi-1D flows, that employs adiabatic and isentropic regimes. 
 The equation of state $p=\rho R T$ essentially rules the fluid stream, where $p$, $\rho$, $T$, and $R$ are  standard notations for the fluid pressure and density, the temperature, and the universal gas constant. The  specific heat ratio shall be denoted by $\gamma$ in what follows and the phenomenological value $\gamma = 1.4$  is adopted. In fact, the air is primarily constituted by diatomic gases, being experimentally consistent with adiabatic indices for dry air in the range 0-300 $\mathring{}$C.
     
 Isentropic fluid streams flow from an initial state to a final one according to the prescription  \cite{landau1987fluid} $p = \rho^\gamma=T^{\frac{\gamma}{\gamma-1}}$. Isentropic fluid flows 
are best used in experiments involving de Laval nozzles, since the flow has a  continuous and  uniform expansion in the nozzle, free of shock waves. The Mach number, 
$
\mathbbmtt{M}\,(x)= \frac{v(x)}{c_s(x)},
$ where $c_s(x)=\sqrt{\frac{dp}{d\rho}}\big\vert_x=\sqrt{\gamma RT(x)}$ denotes the (local) speed of sound, for $x$ standing for the transversal coordinate along the nozzle, and $v$, as usual, denotes the local fluid flow speed. 
Quasi-1D fluid flows are governed by  the conservation laws in hydrodynamics,  \cite{landau1987fluid}, where hereon the notation $(\;\;)_t=\frac{\partial}{\partial t}, (\;\;)_x=\frac{\partial}{\partial x}$ shall be alternatively used:
 \begin{subequations}
\begin{eqnarray}
(\rho A)_t + (\rho Av)_x &\!=\!& 0 \,,
\label{sw0} \\
(\rho Av)_t + [(p+\rho v^2)A]_x &\!=\!& 0 \,,\label{sw1}
\\
\!\!\!\!\!\!\!\!\!\!\!\!\!\!\!\!\!\!\!\!\left(\frac{pA}{1\!-\!\gamma}\!-\!\frac{\rho v^2}{2}\right)_t \!+\!  \left( \frac{\gamma vA}{1\!-\!\gamma}-\frac{\rho v^2}{2} \!\right)_x &\!=\!& 0 \,.\label{sw}
\end{eqnarray}
\end{subequations}
Eq.~\eqref{sw1} is usually replaced by the Euler equation\begin{gather}
\rho(v_t + vv_x)+p_x =0\,,
\label{eueu}
\end{gather}
or equivalently to the Bernoulli one   
\beq\label{enta}
\frac{1}{2}\Phi_x^2 + \int\rho^{-1}dp =-\Phi_t\,,
\label{bno}
\eeq
for $d\Phi/dx = v$.  Eq. (\ref{enta}) yields a linearized equation 
ruling  sound waves. For it, perturbations of the velocity potential and the fluid density are regarded
\beq
\phi=\Phi-\Phi_0,\\
\delta\rho=\rho-\rho_0,\eeq
around  background fields $\Phi_0$ and $\rho_0$  \cite{Okuzumi:2007hf,Abdalla:2007dz}. 
Fluid flows have a stagnation state, with $c_{s0}$ its stagnation speed of sound.
 
Taking the acoustic version of the tortoise coordinate,   $
x^{\star} = c_{s0} \int \left[{ (1-\mathbbmtt{M}^{\,2}(x))c_{s}(x)}\right]^{-1}\!dx 
$, the system (\ref{sw0} -- \ref{sw}) implies that   \cite{Okuzumi:2007hf}
\begin{eqnarray}
\biggl[ \partial_{x_\star}^{2} + \omega^2_0 - V(x_\star) \biggr] \upphi(\omega,x_\star) = 0, \label{seq1},\end{eqnarray} for $\frac{\omega^2}{c_{s0}^2}$, with 
potential \begin{eqnarray}\label{vxx}
V(x_\star) = \frac{1}{2}\left(\frac1{\rm g}\partial^2_{x_{\star}}{\rm g}
    - \frac1{2{\rm g}^2}(\partial_{x_\star}{\rm g})^2\right),
\end{eqnarray}
for \cite{Okuzumi:2007hf,Abdalla:2007dz}
\begin{eqnarray}\label{isen1}
\!\!\!\!\!\!\!\!\!{\rm g}(x) &=&\frac{ A(x)\rho(x)}{c_s(x)},
%\appropto\frac{A(x)}{2\rho^{(\gamma-3)/2}}, 
\\\label{isen11}
\!\!\!\!\!\!\!\!\!\!\upphi(\omega,x_\star) &=\!& \int_{-\infty}^{+\infty} \!\!\!\! e^{i\omega t}\phi(t,x_\star)\,e^{-i\omega f(x_\star)}\sqrt{{\rm g}(x_\star)}\,dt,
\end{eqnarray}
where $f(x)=\int\left(\frac{c_s^2(x)}{v^2(x)}-1\right)^{-1}dx$, in Eq. (\ref{isen11}). 
Here $A^2\appropto(1-\rho^{(\gamma-1)})\rho^2$, what makes Eq. (\ref{isen1}) to yield  \cite{Abdalla:2007dz,Cuyubamba:2013iwa,daRocha:2017lqj}
$
{\rm g}^2=\frac{\rho^{1-\gamma}}{2\left(\rho^{1-\gamma}-1\right)}
$, following that \begin{eqnarray}
\rho^{1-\gamma}&=&G(x)
\label{isen3}
=\frac{\gamma-1}{2}\mathbbmtt{M}\,^2+1,
\end{eqnarray} for 
\beq
\label{GGG}
G(x)\equiv2{\rm g}^2-2{\rm g}\sqrt{{\rm g}^{2}-1}.\eeq It thus implies that 
$\mathbbmtt{M}\,^2=\frac{2}{\gamma-1}(G(x)-1)$. Since the Mach number must be equal to one at the event horizon, the scalar field ${\rm g}(x)$ must be also finite at the horizon, with value 
\begin{eqnarray}\label{condt}
{\rm g}_{\rm horizon}=\frac{1+\gamma}{\gamma-1}\frac1{2\sqrt{2}}\geq 1.\end{eqnarray} 
Eq. (\ref{isen3}) can be substituted into (\ref{isen1}) and the nozzle  
cross-section can be then derived \cite{Okuzumi:2007hf}, 
\begin{eqnarray}\label{2}
A(x)&=&\left(\frac{2}{1+\gamma}-\frac{1-\gamma}{1+\gamma}\mathbbmtt{M}\,^2(x)\right)^{\frac{1-\gamma}{1+\gamma}}\frac{1}{\mathbbmtt{M}\,^2(x)}\,.\label{cross1}
\end{eqnarray}

Fluid flows jets, in de Laval nozzles, were proposed to be a phenomenon that is similar to the ringing of black hole mergers. Indeed, perturbations of scalar fields, as scalar modes describing black hole  backgrounds, are also governed by wave equations with an effective  potential  \cite{Abdalla:2007dz},  
\begin{equation}\label{seq2}
\left(\partial_{r_*}^2+\omega^2-V(r_*)\right)\Psi(r_*)=0,
\end{equation}
where $dr_*=e^{-2\beta(r)}{dr}$, using Eq. \eqref{pert-sol-1th} \cite{Silveira:2017ijo}. The  potential in Eq. (\ref{seq2}) has the form  
\begin{equation}
\!\!\!\!V(r)\!=-\left[\frac1r e^{-4\beta}\left(\beta'-\frac{1}{4r}\right)\right]\bigg\vert_{r=r(r_\ast)}
\end{equation}
Since Eqs. (\ref{seq1}) and (\ref{seq2}) are similar, when an appropriate scalar field  $g(x)$ is chosen in Eq. (\ref{vxx}), the de Laval nozzle is made a dual object to the black hole when both  tortoise coordinates are identified, $dx_\star=dr_*$, implying that 
\begin{eqnarray}
\!\!\!\!\!\!\!\!\!\!\!\!dx_\star^2=
\frac{G(x)\!-\!1}{\left[1\!-\!\frac{2}{\gamma-1}
\left(G(x)\!-\!1\right)\!-\!1\right]^2}dx^2,\label{dxxx}
\end{eqnarray}
or equivalently $dx_\star=\frac{\rho^{{1-\gamma}/2}}{1\!-\!\mathbbmtt{M}\,^2}dx$. 
Hence, the ODE for ${\rm g}(r)$ is obtained, taking into account Eq. \eqref{pert-sol-1th}:
\begin{equation}\label{eq123}
\!\!\!\!\!\!\left[e^{-2\beta}{\rm g}'(r)\right]^\prime\!+2\beta e^{-4\beta}{\rm g}'\!(r)\!-\!\frac{e^{-4\beta}{\rm g}^{\prime2}(r)}{2{\rm g}(r)}\!=\!V(r){\rm g}(r).
\end{equation}
The solution of Eq. \eqref{pert-sol-1th}  can be split into the sum of a pure GR field,  given by ${\rm g}_0(r)\equiv\lim_{\zeta,\tilde\Lambda\to0}{\rm g}(r)$, and a $R^2$ field, as 
\begin{eqnarray}{\rm g}(r)={\rm g}_0(r) + {\rm g}_{R^2},\label{B5B}\end{eqnarray} Hence, one 
can replace the already known solution of Eq. (\ref{eq123}), when  
for $\zeta\to0$ and $\tilde\Lambda\to0$, yielding  \cite{Abdalla:2007dz}
\begin{eqnarray}\nonumber\label{B4B}
\!{\rm g}_0(r)\!=\!{\frac{\gamma+1}{\sqrt{2\gamma\!-\!2}}}\sum_{k=s}^{\ell}\!
\frac12\left[\frac{ (k+\ell)!\,r^{k+1}}{( k + s )! (k-s)! (\ell \!-\! k)!}\!\right]^{\!2}\!.
\end{eqnarray} This expression is the 0$^{\rm th}$ term for iteration to generate 
the ${\rm g}_{R^2}(r)$ as a solution of Eq. (\ref{eq123}), whose two constants of integration  are driven by Eq. (\ref{condt}). Consequently, the de Laval cross-sectional nozzle coordinate $x$ can be derived from (\ref{dxxx}) as a function of the black hole radial coordinate  $r$, using Eq. (\ref{GGG}), as 
\begin{equation}\label{b6b}
\!\!\!x=\intop_{1}^r\!\!\frac{2\left[G(\mathbbmtt{r})-1)\right]-1-\gamma}
{(1\!-\!\gamma)B(\mathbbmtt{r})\left(G(\mathbbmtt{r})-1\right)^{1/2}}\,d\mathbbmtt{r},
\end{equation}
taking into account that  $x=0$ precisely when the fluid flow reaches the speed of sound, namely, at the acoustic event horizon for the sound waves in the fluid \cite{Anacleto:2010cr}.

The nozzle cross-section $A(x)$ is then computed when we replace  Eq. (\ref{B4B}) 
 into  Eq. (\ref{eq123}), using the metric Eq. \eqref{pert-sol-1th}. Numerical analysis is used to compute the scalar field ${\rm g}(r)$  in Eq. (\ref{B5B}), which is expressed in terms of the cross-sectional coordinate $x$, defined  in Eq. (\ref{b6b}).  Since the black hole metric solution with quadratic curvature term, Eq. \eqref{pert-sol-1th}, involves two parameters, 
we can derive bounds on its parameters.
 Straightforward computations, using similar methods as the ones employed in Ref. \cite{Casadio:2015jva}, can use the classical tests of GR to bound those  parameters in Eq. \eqref{pert-sol-1th}. 
The perihelion precession of Mercury yields $|\zeta C_{12}| < (4.3 \pm 3.1) \times 10^{28}$ Kg and the deflection of light by the Sun provides the bound $|\zeta C_{12}| < (7.8 \pm 6.7) \times 10^{29}$ Kg. On the other hand, the gravitational redshift yields $|\zeta C_{12}| <  (8.8 \pm 3.4) \times 10^{29}$ Kg, whereas the radar echo delay finally leads to the bound $|\zeta C_{12}| < (5.1 \pm 4.3) \times 10^{32}$ Kg. We  shall use the most strict bound $|C_{12}| < (4.3 \pm 3.1) \times 10^{28}\,|\zeta^{-1}|$ Kg, to derive the 
nozzle profile. 
 \begin{figure}[H]
\centering\includegraphics[width=8.7cm]{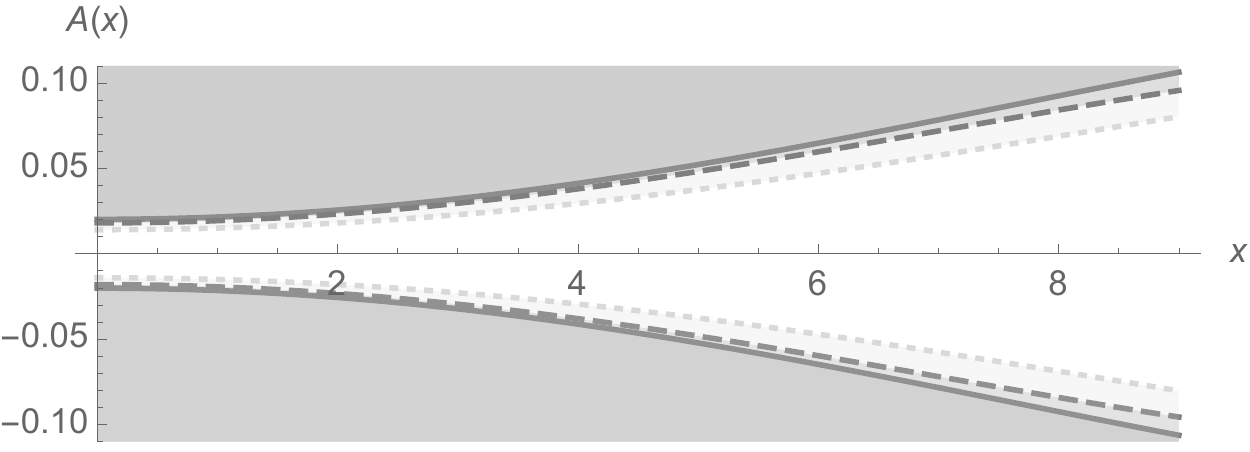}
\caption{The nozzle cross-section $A(x)$  as a function of the cross-sectional nozzle coordinate $x$, for $s=\ell=0$, respectively for the GR limit $\zeta\to 0$ (continuous black line); for $\zeta = 10^{-2}$ (gray dashed line) and for $\zeta = 10^{-1}$ (dotted light gray line). } \end{figure}
 \begin{figure}[H]
\centering\includegraphics[width=8.7cm]{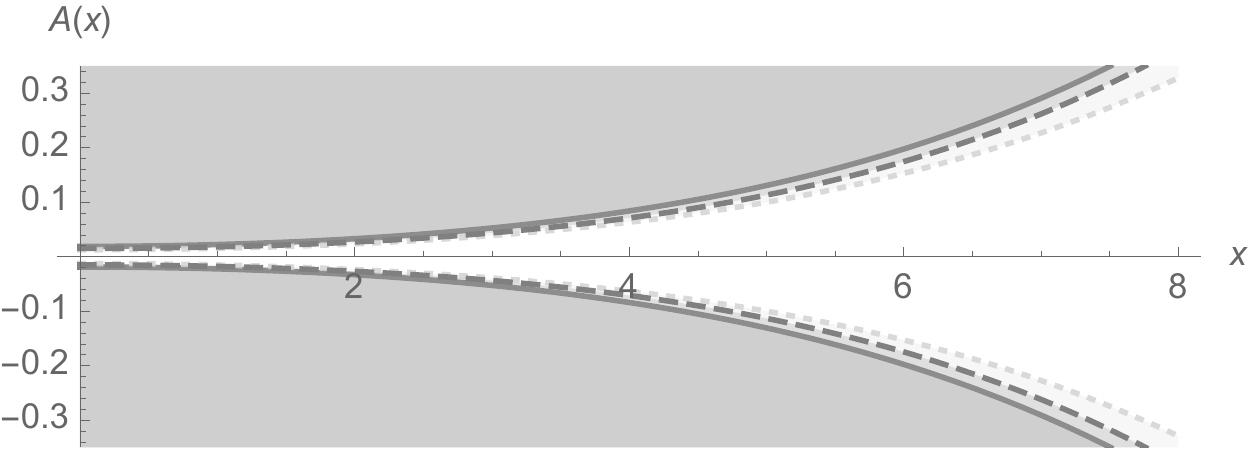}
\caption{The nozzle cross-section $A(x)$  as a function of the cross-sectional nozzle coordinate $x$, for $s=\ell=0$, respectively for the GR limit $\zeta\to 0$ (continuous black line); for $\zeta = 10^{-2}$ (gray dashed line) and for $\zeta = 10^{-1}$ (dotted light gray line). } \end{figure}
It is worth to mention that a higher order term, as the quadratic curvature one in (\ref{ym-map-grav-obs}), beyond the standard Einstein-Hilbert term, provides imprints that can be probed by experiments involving de Laval nozzles in a laboratory, as shown in Fig. 1 and 2. {It then provides  
different signatures  in sonic waves experiments in a de Laval nozzle, corresponding to quasinormal modes emitted by mergers of  black holes  \eqref{pert-sol-1th}}.

\section{Conclusions}
A de Laval propelling nozzle can be constructed upon  black holes that are physical solutions of quadratic curvature gravity, whose metric components are \eqref{pert-sol-1th}. This apparatus has an acoustic event horizon and can produce quasinormal modes of black hole mergers in a propulsion laboratory, probing higher order curvature terms in theories of gravity. 
The corrections to the Schwarzschild black hole solution, observed in Figs. 1 and 2, arise due to the  different fluid pressure regime across the nozzle.  In those figures,  the most strict bound $|C_{12}| < (4.3 \pm 3.1) \times 10^{28}\,|\zeta^{-1}|$ Kg, provided by the classical tests of GR for perihelion precession of Mercury case, was employed used to derive the de Laval nozzle profile. The bigger the parameter $\zeta$, that drives the $R^2$ corrections in the metric \eqref{pert-sol-1th}, the smaller the nozzle cross-sectional area is. It  shows that the $R^2$ corrections decrease the nozzle cross-sectional area, yielding, by Eq. (\ref{2}), a bigger Mach number. Hence, the fluid flow speed increases and the sound speed decreases, instead. These results imply a sonic point corresponding to a lower speed of sound and, consequently, a modified event horizon for the sound waves through the nozzle. Moreover, supersonic regions are then reached with lower flow speeds,  corresponding to the sonic black hole inner region.

\subsection*{Acknowledgements}

RdR~is grateful to CNPq (Grant No. 303293/2015-2),
 to FAPESP (Grant No.~2015/10270-0), for partial financial support.
RS and ATT thank to CNPq-Brazil, CAPES, PROPPI-UFF and CBPF for financial support. 

\bibliographystyle{iopart-num}
\bibliography{R2deLavalfim}

\end{document}